\documentclass[twocolumn,aps,prc,showpacs,floatfix]{revtex4}
\usepackage{graphicx}
\usepackage{dcolumn}
\usepackage{bm}

\begin{document}

\title{Effects of elastic and inelastic $NN$ scattering cross
sections on $\pi^{-}/\pi ^{+}$ ratios in heavy-ion collisions at
intermediate energies}
\author{Gao-Chan Yong}
\affiliation{Institute of Modern Physics, Chinese Academy of
Sciences, Lanzhou 730000, China}

\begin{abstract}
Based on the isospin-dependent Boltzmann-Uehling-Uhlenbeck
transport model and the scaling model according to nucleon
effective mass, effects of elastic and inelastic $NN$ scattering
cross sections on $\pi^{-}/\pi ^{+}$ in the neutron-rich reaction
of $^{48}$Ca+$^{48}$Ca at a beam energy of $400$ MeV/nucleon are
studied. It is found that cross-section effects of both $NN$
elastic and inelastic scatterings affect $\Delta_{1232}$,
$\pi^{-}$ and $\pi^{+}$ production, as well as the value of
$\pi^{-}/\pi^{+}$.
\end{abstract}

 \maketitle


Recently, pion production in heavy-ion collisions has attracted
much attention in the nuclear physics community
\cite{xiao09,ditoro1,xu09,Rei07,yong06}. One important reason for
this is that pion production is connected with the high-density
behavior of nuclear symmetry energy \cite{LiBA02}. The latter is
crucial for understanding many interesting issues in both nuclear
physics and astrophysics
\cite{Bro00,Dan02a,Bar05,LCK08,Sum94,Lat04,Ste05a}. The high-%
density behavior of nuclear symmetry energy, however, has been
regarded as the most uncertain property of dense neutron-rich
nuclear matter \cite{Kut94,Kub99}. Many microscopic and/or
phenomenological many-body theories using various interactions
\cite{Che07,LiZH06} predict that the symmetry energy increases
continuously at all densities. However, other models
\cite{Pan72,Fri81,Wir88a,Kra06,Szm06,Bro00,Cha97,Sto03,Che05b,Dec80,MS,Kho96,Bas07,Ban00}
predict that the symmetry energy first increases to a maximum and
then may start decreasing at certain suprasaturation densities.
Thus, currently the theoretical predictions on the symmetry energy
at supra-saturation densities are extremely diverse. To make
further progress in determining the symmetry energy at
suprasaturation densities, what is most critically needed is some
guidance from dialogues between experiments and transport models,
which have been done extensively in the studies of nuclear
symmetry energy at low densities
\cite{tsang09,shetty07,fami06,tsang04,chen05}.

Using $\pi^-/\pi^+$ to probe the high-density behavior of nuclear
symmetry energy has evident advantage within both the $\Delta$
resonance model and the statistical model \cite{Sto86,Ber80}.
Several hadronic transport models have quantitatively shown that
$\pi^-/\pi^+$ ratio is indeed sensitive to the symmetry energy
\cite{LiBA02,yong06,Gai04,LiQF05b}, especially around pion
production threshold. These transport models, however, usually use
different elastic and inelastic $NN$ scattering cross sections.
For the in-medium $NN$ elastic scattering cross section, different
transport models use different forms. For the $NN$ inelastic
scattering cross section, they usually use free $NN$ inelastic
scattering cross section. The in-medium $NN$ inelastic scattering
cross section must be different from that in free space, and
currently the in-medium $NN$ inelastic scattering cross section is
quite controversial \cite{ditoro2,Lar01,Lar03,Ber88,Mao97}. More
importantly, the description of meson production in heavy-ion
collisions is still a very open problem. Other effects, apart from
the one considered here, may change the $\pi^-/\pi^+$ ratio. For
instance, in Ref. \cite{xu09}, the authors discuss in-medium
effects, owing to the interaction of pions with nucleons, on the
charged-pion ratio, which go in the direction of reducing
$\pi^-/\pi^+$ ratio. However, the cross-section reduction effects
investigated in the following increase the charged-pion ratio. All
these effects on pion production are combined with the effects of
the isovector part of the nuclear interaction (the symmetry
energy), which influences strongly also the isotopic content of
pre-equilibrium nucleon emission, changing the asymmetry of the
remaining interacting system. From this point of view, it is not
so easy to extract information on the high-density behavior of the
symmetry energy just looking at pion production and charged-pion
ratio, because several effects cooperate to build the final
result. All the preceding may cause different translations from
experimental data \cite{Lar01,Lar03,factor,Liqelast06}. Here we
just study the effects of both elastic and inelastic $NN$
scattering
cross sections on pion production, as well as the value of $\pi^{-}/\pi%
^{+}$ in neutron-rich heavy-ion collisions because the National
Superconducting Cyclotron Laboratory at Michigan State University,
Rikagaku Kenkyusho (RIKEN, The Institute of Physical and Chemical
Research) of Japan, and the Cooler Storage Ring in Lanzhou, China,
are planning to do experiments of pion production to study the
high-density behavior of nuclear symmetry energy. In the framework
of the isospin-dependent Boltzmann-Uehling- Uhlenbeck (IBUU)
transport model, as an example, we studied the effects of both
elastic and inelastic $NN$ scattering cross sections on
$\pi^{-}/\pi ^{+}$ in the neutron-rich reaction of
$^{48}$Ca+$^{48}$Ca at a beam energy of $400$ MeV/nucleon. It is
found that cross-section effects of both $NN$ elastic and
inelastic scatterings affect $\Delta_{1232}$, $\pi^{-}$ and
$\pi^{+}$ production, as well as the value of $\pi^{-}/\pi^{+}$.


The isospin and momentum-dependent mean-field potential used in
the present work is \cite{Das03}
\begin{eqnarray}
U(\rho, \delta, \textbf{p},\tau)
=A_u(x)\frac{\rho_{\tau^\prime}}{\rho_0}+A_l(x)\frac{\rho_{\tau}}{\rho_0}\nonumber\\
+B\left(\frac{\rho}{\rho_0}\right)^\sigma\left(1-x\delta^2\right)\nonumber
-8x\tau\frac{B}{\sigma+1}\frac{\rho^{\sigma-1}}{\rho_0^\sigma}\delta\rho_{\tau^{\prime}}\nonumber\\
+\sum_{t=\tau,\tau^{\prime}}\frac{2C_{\tau,t}}{\rho_0}\int{d^3\textbf{p}^{\prime}\frac{f_{t}(\textbf{r},
\textbf{p}^{\prime})}{1+\left(\textbf{p}-
\textbf{p}^{\prime}\right)^2/\Lambda^2}},
\label{Un}
\end{eqnarray}
where $\rho_n$ and $\rho_p$ denote neutron ($\tau=1/2$) and proton
($\tau=-1/2$) densities, respectively.
$\delta=(\rho_n-\rho_p)/(\rho_n+\rho_p)$ is the isospin asymmetry
of nuclear medium. All parameters in the preceding equation can be
found in refs. \cite{IBUU04}. The variable $x$ is introduced to
mimic different forms of the symmetry energy predicted by various
many-body theories without changing any property of symmetric
nuclear matter and the value of symmetry energy at normal density
$\rho_0$. In this article we let the variable $x$ be $1$. With
these choices the symmetry energy obtained from the preceding
single-particle potential is consistent with the Hartree-Fock
prediction using the original Gogny force \cite{Das03} and is also
favored by recent studies based on FOPI experimental data
\cite{xiao09}. The main reaction channels related to pion
production and absorption are
\begin{eqnarray}
&& NN \longrightarrow NN, \nonumber\\ && NR \longrightarrow NR,
\nonumber\\ && NN \longleftrightarrow NR, \nonumber\\ && R
\longleftrightarrow N\pi,
\end{eqnarray}
where $R$ denotes $\Delta $ or $N^{\ast }$ resonances. In the
present work, we use the isospin-dependent in-medium reduced $NN$
elastic scattering cross section from the scaling model according
to nucleon effective mass \cite{factor,neg,pan,gale} and compare
with the case of $NN$ elastic cross section in free space to study
the effect of elastic $NN$ scattering cross section on pion
production. Assuming in-medium $NN$ scattering transition matrix
is the same as that in vacuum \cite{pan}, the elastic $NN$
scattering cross section in medium $\sigma _{NN}^{medium}$ is
reduced compared with their free-space value $\sigma _{NN}^{free}$
by a factor of
\begin{eqnarray}
R_{medium}(\rho,\delta,\textbf{p})&\equiv& \sigma
_{NN_{elastic}}^{medium}/\sigma
_{NN_{elastic}}^{free}\nonumber\\
&=&(\mu _{NN}^{\ast }/\mu _{NN})^{2}.
\end{eqnarray}
where $\mu _{NN}$ and $\mu _{NN}^{\ast }$ are the reduced masses
of the colliding nucleon pair in free space and medium,
respectively. For in-medium $NN$ inelastic scattering cross
section, even assuming in-medium $NN \rightarrow NR$ scattering
transition matrix is the same as that in vacuum, the density of
final states $D_{f}^{'}$ \cite{pan} of $NR$ is very hard to
calculate owing to the fact that the resonance's potential in
matter is presently unknown. Because the purpose of present work
is just study the effect of $NN$ scattering cross section on pion
production and charged-pion ratio, to simplify the question, for
the $NN$ inelastic scattering cross section we use the same
correction factor $R_{medium}(\rho,\delta,\textbf{p})$ to study
the effect of $NN$ inelastic scattering cross sections on pion
production and charged-pion ratio (i.e., one choice is
$R_{medium}(\rho,\delta,\textbf{p})\ast\sigma
_{NN_{inelastic}}^{free}$, the other choice is $\sigma
_{NN_{inelastic}}^{free}$). The effective mass of nucleon in
isospin asymmetric nuclear matter is
\begin{equation}
\frac{m_{\tau }^{\ast }}{m_{\tau }}=\left\{ 1+\frac{m_{\tau }}{p}\frac{%
dU_{\tau }}{dp}\right\}^{-1}.
\end{equation}
From the definition and Eq.~(\ref{Un}), we can see that the
effective mass depends not only on density and asymmetry of medium
but also the momentum of nucleon. We decide $NR\rightarrow NN$
inelastic scattering cross section according to the detailed
balance principle \cite{Lar03}. The reduction factor $R_{medium}$
thus affects not only pion production but also pion absorption.
\begin{figure}[th]
\begin{center}
\includegraphics[width=0.5\textwidth]{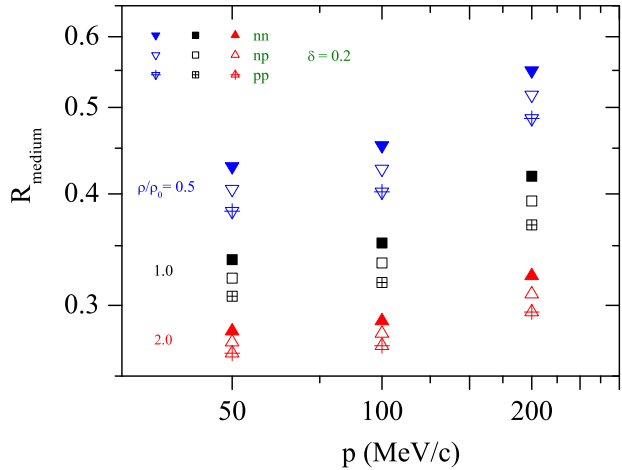}
\end{center}
\caption{The reduction factor $R_{medium}$ as a function of
density (0.5, 1.0, and 2.0 times normal nuclear matter density
$\rho_{0}$) and momentum (50, 100 and 200 MeV/c) for $nn$
(neutron-neutron), $np$ (neutron-proton), and $pp$ (proton-proton)
colliding nucleon pairs. The asymmetry of nuclear matter is set to
be $\delta=0.2$.} \label{factor}
\end{figure}
\begin{figure}[th]
\begin{center}
\includegraphics[width=0.5\textwidth]{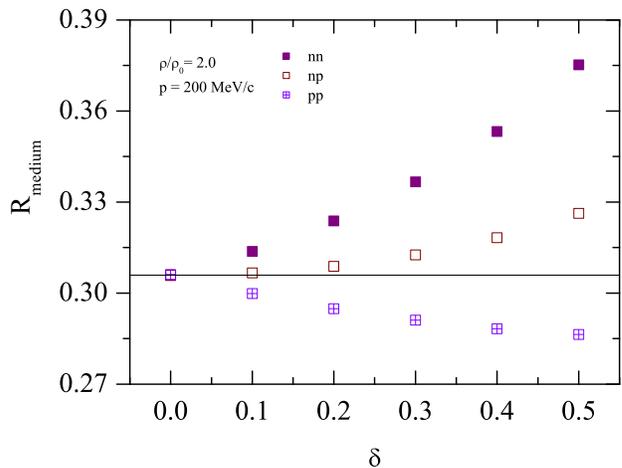}
\end{center}
\caption{The reduction factor $R_{medium}$ as a function of
asymmetry $\delta$ of medium for $nn$, $np$, and $pp$ colliding
nucleon pairs. The density of medium is $\rho/\rho_{0}=2.0$ and
nucleonic momentum is $p=200MeV/c$.} \label{factor2}
\end{figure}
Shown in Fig.~\ref{factor} is the reduction factor $R_{medium}$ as
a function of density and momentum for $nn$ (neutron-neutron),
$np$ (neutron-proton), and $pp$ (proton-proton) colliding nucleon
pairs. We can see that the reduction factor decreases with density
and increases with momentum. Also, we can clearly see that the
$nn$ pair's reduction factor is always larger than that of the
$pp$ pair in neutron-rich nuclear matter. Fig.~\ref{factor2} shows
the reduction factor $R_{medium}$ as a function of the asymmetry
$\delta$ of the medium. It is seen that the reduction factor
$R_{medium}$ of the $nn$ colliding nucleon pair increases rapidly
with asymmetry $\delta$ and the reduction factor $R_{medium}$ of
the $np$ colliding nucleon pair increases slowly with asymmetry
$\delta$, whereas the reduction factor $R_{medium}$ of the $pp$
colliding nucleon pair decreases slowly with asymmetry $\delta$.
In the symmetric nuclear matter ($\delta$= 0), however, the
reduction factor $R_{medium}$ for the three cases does not split.


\begin{figure}[th]
\begin{center}
\includegraphics[width=0.5\textwidth]{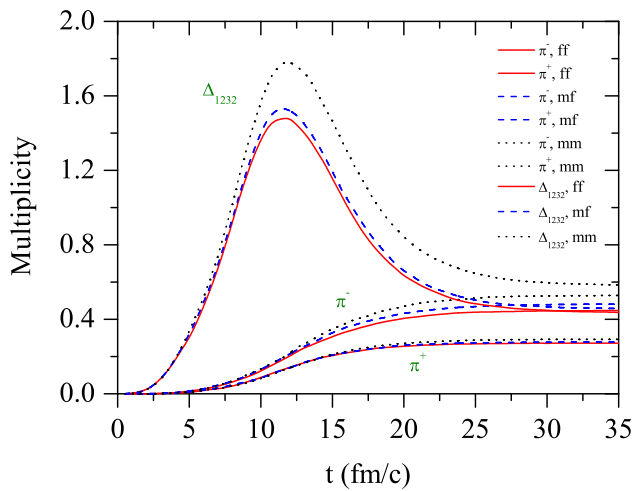}
\end{center}
\caption{Evolution of $\pi^{-}$, $\pi ^{+}$ and $\Delta(1232)$
multiplicities in the reaction of $^{48}$Ca+$^{48}$Ca at a beam
energy of 400 MeV/nucleon. The ``ff'', ``mf'' and ``mm'' denote
the $NN$ free elastic plus $NN$ free inelastic scattering cross
sections, modified $NN$ reduced elastic plus $NN$ free inelastic
scattering cross sections and modified $NN$ reduced elastic plus
modified $NN$ reduced inelastic scattering cross sections,
respectively.} \label{multi}
\end{figure}
To study the effects of both $NN$ elastic and inelastic scattering
cross sections on $\pi^{-}/\pi ^{+}$, We first studied their
effects on $\Delta(1232)$ and pion productions. We ignore
$N^{\ast}$ production here, because it is mainly related to more
energetic collisions. Shown in Fig.~\ref{multi} is the evolution
of pion and $\Delta(1232)$ multiplicities in the reaction of
$^{48}$Ca+$^{48}$Ca at a beam energy of 400 MeV/nucleon. The
``ff'', ``mf'' and ``mm'' denote the $NN$ free elastic plus $NN$
free inelastic scattering cross sections (i.e., full free $NN$
scattering cross sections), modified $NN$ reduced elastic plus
$NN$ free inelastic scattering cross sections and modified $NN$
reduced elastic plus modified $NN$ reduced inelastic scattering
cross sections (i.e., full modified $NN$ reduced scattering cross
sections), respectively. First, we can see that $\Delta(1232)$
production is mainly at the compression stage. Compared with the
full free $NN$ scattering cross-sections case, both the modified
$NN$ reduced elastic and the modified $NN$ reduced inelastic
scattering cross sections cause more $\Delta(1232)$ resonance
production. This is because for the ``mf'' case, the modified $NN$
reduced elastic scattering cross section causes the colliding
nuclei to show less stopping \cite{Liqelast06,Liu01,Lar01}; as a
result more energetic $NN$ collisions in fireball matter produce
more resonances. For the ``mm'' case, the reduction factor
$R_{medium}$ of the $NN$ inelastic scattering cross section not
only makes the colliding nuclei further less stopping but also
makes resonance less absorptive (i.e., the cross section of
$NR\rightarrow NN$ decreases). Effects of the preceding two
factors are larger than that of the modified $NN$ reduced
inelastic cross section. Therefore, number of resonances also
relatively increase. More resonances produce more $\pi^{-}$'s and
$\pi^{+}$'s. We thus see more $\pi^{-}$'s and $\pi^{+}$'s for the
``mm'' case than for the ``mf'' case and than for the ``ff'' case.
Second, one can see that $\pi^{-}$ production is more sensitive to
$NN$ scattering cross sections than $\pi^{+}$. It is known that
$nn$ collisions mainly produce $\pi^{-}$, $pp$ collisions mainly
produce $\pi^{+}$ and $np$ collisions produce roughly equal
numbers of $\pi^{-}$ and $\pi^{-}$ \cite{Sto86}. Although the
modified $NN$ reduced cross section makes the colliding nuclei
less stopping, relative to $nn$ collisions, the Coulomb force
between proton and proton almost cancel out the effect of a
largely modified $pp$ reduced cross section (shown in
Fig.~\ref{factor2}). Therefore $\pi^{-}$ production is more
sensitive to the $NN$ elastic and inelastic scattering cross
sections than $\pi^{+}$. Finally, we can clearly see that more
$\pi^{-}$'s are produced than $\pi^{+}$'s. This is understandable
for neutron-rich $^{48}$Ca+$^{48}$Ca collision, in which there are
more $nn$ collisions than $pp$ collisions \cite{LYZ}.

\begin{figure}[th]
\begin{center}
\includegraphics[width=0.5\textwidth]{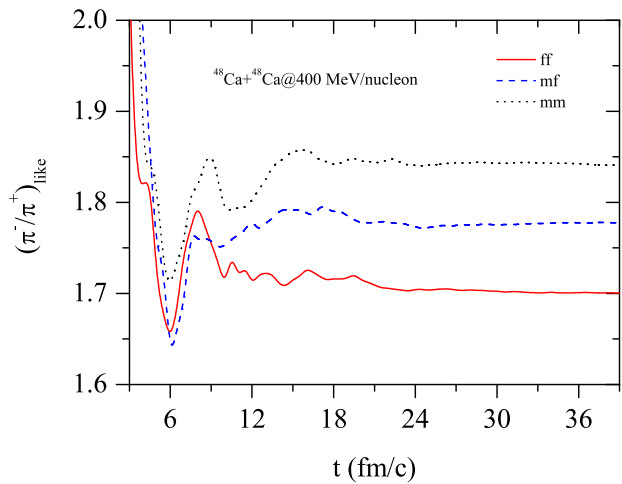}
\end{center}
\caption{Effects of both elastic and inelastic $NN$ scattering
cross sections on $(\pi^-/\pi^+)_{like}$ as a function of time in
the reaction of $^{48}$Ca+$^{48}$Ca at a beam energy of $400$
MeV/nucleon.} \label{medRpion}
\end{figure}
To reduce the systematic errors in simulations, especially in
experimental analysis, one usually studies the $\pi^{-}/\pi ^{+}$
\cite{yong06,LYZ,Rei07} instead of $\pi^{-}$ or $\pi^{+}$ only.
Shown in Fig.~\ref{medRpion} are effects of elastic and inelastic
$NN$ scattering cross sections on the $(\pi^-/\pi^+)_{like}$ as a
function of time in the central reaction of $^{48}$Ca+$^{48}$Ca at
a beam energy of $400$ MeV/nucleon. In the dynamics of pion
resonance productions and decays the $(\pi^-/\pi^+)_{like}$ reads
\cite{LYZ}
\begin{equation}
(\pi^-/\pi^+)_{like}\equiv
\frac{\pi^-+\Delta^-+\frac{1}{3}\Delta^0}
{\pi^++\Delta^{++}+\frac{1}{3}\Delta^+}.
\end{equation}
This ratio naturally becomes the $\pi^-/\pi^+$ ratio at the
freeze-out stage \cite{LYZ}. From Fig.~\ref{medRpion} we can first
see that, owing to the abundant neutron-neutron scatterings when
the two neutron skins start overlapping at the beginning of the
reaction, the $(\pi^-/\pi^+)_{like}$ ratio reaches a very high
value in the early stage of the reaction. The sensitivity of
$(\pi^-/\pi^+)_{like}$ to the effects of both elastic and
inelastic $NN$ scattering cross sections is clearly shown after
$t=15 fm/c$. We can see that the $\pi^-/\pi^+$ value using full
free $NN$ scattering cross sections is smaller than that of using
modified $NN$ reduced elastic scattering cross section. The
$\pi^-/\pi^+$ value using modified $NN$ reduced elastic scattering
cross section is also smaller than that of using full modified
$NN$ reduced scattering cross sections. This is understandable
from the preceding charged-pion production analysis shown in
Fig.~\ref{multi}. We also did a calculation for central Au+Au
reaction at $400$ MeV/nucleon and found that compared with the
``mf'' case used in Ref.~\cite{xiao09}, with the ``mm'' case the
$\pi^-/\pi^+$ ratio increases about $7\%$ while the charged-pion
multiplicity increases about $30\%$. Hence the effects of $NN$
elastic and inelstic scattering cross sections play an important
role in studying the high-density behavior of nuclear symmetry
energy by using $\pi^-/\pi^+$ in \emph{neutron-rich} heavy-ion
collisions.


In conclusion, in the framework of the isospin-dependent transport
model IBUU and the scaling model according to nucleon effective
mass, we studied the effects of elastic and inelastic $NN$
scattering cross sections on $\pi^{-}/\pi ^{+}$ in the
neutron-rich reaction of $^{48}$Ca+$^{48}$Ca at a beam energy of
$400$ MeV/nucleon. We find that both $NN$ elastic and inelastic
scattering cross sections in neutron-rich heavy-ion collisions
affect $\Delta_{1232}$, $\pi^{-}$ and $\pi^{+}$ production, as
well as the value of $\pi^{-}/\pi^{+}$. The reduced $NN$ elastic
and inelastic cross sections increase the number of $\pi^{-}$ more
evidently than $\pi^{+}$. The value of $\pi^{-}/\pi ^{+}$ thus
also increases accordingly. It is expected that the choice adopted
for the NN cross section affects significantly also the amount of
pre-equilibrium nucleon emission and the corresponding N/Z ratio,
which is also an observable widely investigated in heavy-ion
reactions, from both the experimental and the theoretical point of
view. Such an observable also needs to be carefully investigated,
in parallel with meson production, before reaching any conclusion
about the cross sections and the behavior of the symmetry energy
at high densities.


This work is supported in part by the National Natural Science
Foundation of China under grants 10740420550, 10875151.

\end{document}